\documentclass[prl,superscriptaddress,longbibliography,footinbib,twocolumn,preprintnumbers,amsmath,amssymb]{revtex4-2}

\usepackage{amssymb}
\usepackage{amsmath}
\usepackage{comment} 

\usepackage{xr}
\makeatletter
\newcommand*{\addFileDependency}[1]{
  \typeout{(#1)}
  \@addtofilelist{#1}
  \IfFileExists{#1}{}{\typeout{No file #1.}}
}
\makeatother

\newcommand*{\myexternaldocument}[1]{%
    \externaldocument{#1}%
    \addFileDependency{#1.tex}%
    \addFileDependency{#1.aux}%
}

\myexternaldocument{supp}

\usepackage{xcolor}
\usepackage[pdftex]{graphicx}

\usepackage{empheq}

\usepackage{mathtools}

\usepackage{makecell}

\usepackage{siunitx}
\ifdefined\unit\else
  \ifdefined\NewCommandCopy
    \NewCommandCopy\unit\si
  \else
    \NewDocumentCommand\unit{O{}m}{\si[#1]{#2}}
  \fi
\fi

\usepackage{enumerate}
\usepackage{mathtools}
\usepackage{stmaryrd}

\usepackage{enumitem}

\usepackage{textcomp}
\usepackage{gensymb}

\newcommand{\ex}[1]{\mathrm{e}^{#1}}

\newcommand{\dd}[0]{\mathrm{d}}

\newcommand{\rr}[0]{\boldsymbol{r}}

\newcommand{\te}[0]{\text{e}}
\newcommand{\ts}[0]{\text{s}}
\newcommand{\tp}[0]{\text{p}}

\usepackage[normalem]{ulem}
\usepackage{cancel}

\definecolor{darkblue}{rgb}{0,0,0.6}
\definecolor{darkred}{rgb}{0.6,0,0}
\usepackage[colorlinks=true,urlcolor=darkblue,citecolor=darkblue,linkcolor=darkred]{hyperref}

\begin{document}

\title{Catalysis-induced phase separation of enzymes \emph{via} diffusivity contrast}

\author{Marie Jardat}
\affiliation{Sorbonne Universit\'e, CNRS, Physical Chemistry of Electrolytes and Interfacial Nanosystems (PHENIX), 4 place Jussieu, Paris, France}

\author{Jacques Fries}
\affiliation{Sorbonne Universit\'e, CNRS, Physical Chemistry of Electrolytes and Interfacial Nanosystems (PHENIX), 4 place Jussieu, Paris, France}

\author{Vincent Dahirel}
\affiliation{Sorbonne Universit\'e, CNRS, Physical Chemistry of Electrolytes and Interfacial Nanosystems (PHENIX), 4 place Jussieu, Paris, France}

\author{Jaime Agudo-Canalejo}
\email{j.agudo-canalejo@ucl.ac.uk}
\affiliation{Department of Physics and Astronomy, University College London, London WC1E 6BT, United Kingdom}

\author{Pierre Illien}
\email{pierre.illien@sorbonne-universite.fr}
\affiliation{Sorbonne Universit\'e, CNRS, Physical Chemistry of Electrolytes and Interfacial Nanosystems (PHENIX), 4 place Jussieu, Paris, France}

\date{\today}

\begin{abstract}
Catalysis can drive phase separation by coupling local chemical composition to molecular transport, even in the absence of attractive interactions. Here, we demonstrate such catalysis-induced phase separation  in a microscopic particle-based model of enzymes catalyzing the interconversion of substrate and product molecules with different diffusivities. We derive an analytical instability criterion in the dilute limit, showing that a homogeneous mixture becomes unstable when the nonequilibrium chemical drive and the diffusivity contrast are sufficiently large, and the substrate and product concentration exceeds a threshold set by the enzyme volume. Brownian-dynamics simulations confirm the predicted instability and reveal macroscopic coexistence of enzyme-rich and enzyme-poor phases, including in regimes where equilibrium depletion interactions cannot account for demixing. Finally, we show that, upon eliminating the fast chemical degrees of freedom, the reactive ternary mixture maps onto an effective binary mixture nonreciprocal interactions. Our results establish an explicit microscopic mechanism for catalysis-induced phase separation  and connect chemically driven phase separation to the physics of motility-regulated and nonreciprocal matter.

\end{abstract}

\maketitle

\emph{Introduction.---} Phase separation is a central organizing principle of living matter, underlying the formation of membraneless compartments such as biomolecular condensates~\cite{Hyman2014,Banani2017}. At equilibrium, demixing requires effective attractions between components, encoded in a free energy. Living systems, however, operate far from equilibrium, which raises the question of whether nonequilibrium activity alone can drive phase separation. In nonreactive active matter, this is the case for self-propelled particles, which undergo motility-induced phase separation~\cite{Cates2015}, and for mixtures of particles coupled to different thermostats~\cite{Grosberg2015,PhysRevLett.116.058301,Ilker2020,Jardat2022}. In chemically active systems, where particles take part in nonequilibrium reactions, theoretical studies have mostly considered reactions that switch particles between states with different interactions~\cite{Glotzer1994,Glotzer1994a,Christensen1996,Weber2019,Kirschbaum2021,Zwicker2022,Moncho-Jorda2020,Bley2021,Decayeux2021,Bley2022,Decayeux2022,Goth2022,Alston2022a,Decayeux2023,Alston2026,Holschuh2025}: activity then controls the size, number and dynamics of condensates, but the driving force for demixing is still due to the modulation of direct, equilibrium-like interactions~\cite{Berthin2025,Fries2025a,Fries2025b}.

\begin{figure}[b]
    \centering
    \includegraphics[width=\columnwidth]{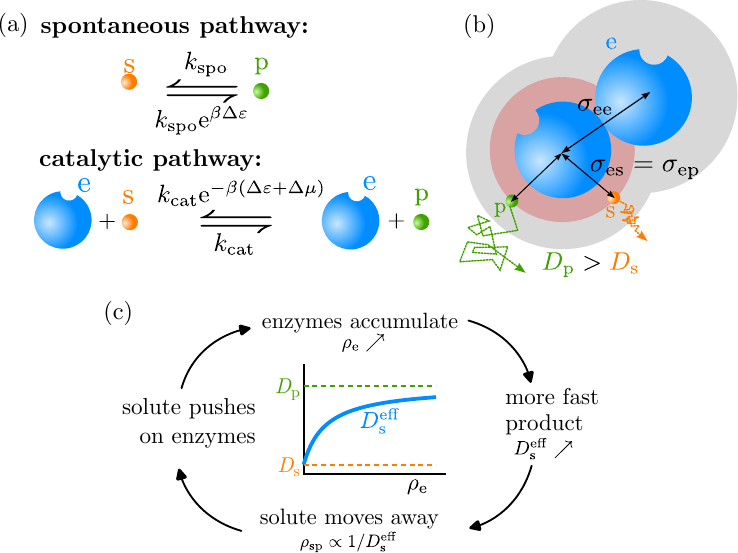}
    \caption{(a)~Spontaneous and catalyzed interconversion of substrate and product. For $\Delta \varepsilon<0$ and $\Delta \varepsilon + \Delta \mu >0$, the spontaneous (resp. catalytic) reaction favors p $\to$ s (resp. s $\to$ p). (b)~Molecules of species $i$ and $j$ cannot come closer to each other than a distance denoted by $\sigma_{ij}$. For catalysis-induced phase separation (CIPS) to occur, product molecules must diffuse faster than substrate molecules ($D_\text{p}>D_\text{s}$). (c)~Mechanism of CIPS, where $D_\ts^\text{eff}$ represents the effective diffusivity of a solute molecule (substrate and product combined).}
    \label{fig:system_sketch}
\end{figure}

A recently proposed mechanism, termed catalysis-induced phase separation (CIPS), suggests a fundamentally different route to nonequilibrium demixing~\cite{Cotton2022}. Rather than modifying particle interactions, fuelled chemical reactions catalyzed by enzymes convert molecules between states with different transport properties, thereby coupling local composition to transport and potentially destabilizing a homogeneous mixture in the absence of any attractive interaction. In Ref.~\cite{Cotton2022}, this mechanism was established within a Flory--Huggins description of an incompressible enzyme--substrate--product mixture, in which the key ingredient is an asymmetry between the off-diagonal Onsager mobilities that couple the enzyme flux to the substrate and product thermodynamic forces ($D_{\tp\te}\neq D_{\ts\te}$). This asymmetry was introduced phenomenologically, and its microscopic origin was left open. Moreover, in the simplest, solvent-free version of the theory, substrate and product molecules fill the whole volume not occupied by enzymes, so that their concentration does not appear as an independent control parameter. A microscopic, particle-based description is therefore needed to identify the minimal physical ingredients required for CIPS, to express its threshold in terms of molecular sizes and concentrations, and to assess how it competes with the equilibrium depletion forces that are unavoidable in mixtures of large enzymes and small solutes~\cite{Oosawa1954,Asakura1958,Binder2014}.

Here we address these questions using a particle-based model of enzymes, substrates, and products, where the latter two undergo fuelled catalytic interconversion.  We derive an analytical instability criterion and confirm it with Brownian-dynamics simulations, which reveal macroscopic demixing. We show that the transport asymmetry required for CIPS emerges from a mere difference between the self-diffusion coefficients of substrate and product, the enzymes interacting identically with both, and that, in the presence of a solvent, CIPS requires a threshold substrate and product concentration. Furthermore, we show that the resulting dynamics maps onto an effective binary mixture in which the solute diffusivity is regulated by the local enzyme density, which yields nonreciprocal enzyme--solute couplings and establishes a direct connection between CIPS, motility-regulated active matter~\cite{Tailleur2008,Cates2015}, and nonreciprocal matter~\cite{Saha2020,You2020,Fruchart2026}.
 Importantly, unlike mixtures of species connected to different thermostats~\cite{Grosberg2015,PhysRevLett.116.058301,Jardat2022,Ilker2020,Ilker2021,Schwarcz2024a, Wang2020a}, our model maintains fluctuation-dissipation with a single thermostat, and the only source of nonequilibrium is the fuelled catalytic reactions.

\emph{Model.---} We consider a three-dimensional, ternary mixture made of enzyme, substrate and product molecules. We assume that all the molecules obey overdamped Langevin (Brownian) dynamics, with bare hydrodynamic mobilities (inverse friction) $\mu_i$ and, via Einstein's relation, diffusion coefficients $D_i=\mu_i k_\mathrm{B} T$, where the index $i\in\{\te,\ts,\tp\}$ denotes  species. Molecules of species $i$ and $j$ interact through a radially symmetric pair potential $V_{ij}(r)$, which we will take to be purely repulsive. We assume that the conversions between substrate and product molecules occur via two pathways [Fig.~\ref{fig:system_sketch}(a)]. First, a \textit{spontaneous} pathway $\text{s} \leftrightharpoons \text{p}$ may occur anywhere in the system, with direct and reverse conversion rates  respectively given by  $k_\text{spo}\ex{\beta\Delta\varepsilon}$ and $k_\text{spo}$, where $\Delta\varepsilon\equiv \varepsilon_\ts-\varepsilon_\tp$ is the difference between the internal energies of substrate and product molecules, and $\beta \equiv (k_\mathrm{B}T)^{-1}$. These rates fulfill detailed balance by construction. Second,  a \textit{catalytic} pathway $\text{e}+\text{s} \leftrightharpoons \text{e}+\text{p}$ only occurs in the vicinity of enzyme molecules, with direct and reverse rates per molecule given by $k_\text{cat}\rho_\te$ and $k_\text{cat}\rho_\te\ex{-\beta(\Delta\varepsilon+\Delta\mu)}$ respectively, where $\rho_\te$ is the local enzyme density (so that $k_\text{cat}$ is a second-order rate constant), where $\Delta \mu$ is a nonequilibrium drive, that represents the possibility for the catalytic reaction to involve additional fuel and waste species, whose chemical potential are maintained constant through the presence of reservoirs. The catalytic pathway thus breaks detailed balance as soon as $\Delta\mu\neq 0$.
To lighten the notation, we will use $\varepsilon \equiv \beta\Delta\varepsilon$ and $\mu \equiv \beta\Delta\mu$ in the following. Finally, we will assume that substrate and product molecules interact in the same way with each other (i.e. $V_{\ts\ts}=V_{\tp\tp}=V_{\ts\tp}$), and with enzymes (i.e. that $V_{\te\ts} = V_{\te\tp}$), but that they generally have different hydrodynamic mobilities and thus different diffusion coefficients ($D_\tp\neq D_\ts$), see Fig.~\ref{fig:system_sketch}(b).

\emph{Low-density stability criterion.---} We first establish the reaction-diffusion equations obeyed by the one-body local densities $\rho_i$. 
Pair interactions between molecules are encoded in the diffusive part of the evolution equations, and generally make the problem intractable (see Supplemental Material~\cite{SM} for a complete formulation in terms of the $N$-body Smoluchowski equation). In the low-density limit, however, analytical progress can be made, and we find that the dynamics can be written as $\partial_t \rho_i = \nabla \cdot \sum_{j}\mathcal{D}_{ij} \nabla \rho_j + R_i$ with the diffusion matrix $\mathcal{D}_{ij}=D_i (\delta_{ij}+\rho_i \mathcal{B}_{ij})$,
where we have introduced the virial coefficients $\mathcal{B}_{ij} \equiv \int \dd \rr [1-\ex{-\beta V_{ij}(\rr)}]$.
The reaction fluxes for each species read $R_\te=0$, $R_{\ts,\tp} = \pm\{	 k_\text{spo} \left[ \rho_\tp-\ex{\varepsilon} \rho_\ts \right]
+ k_\text{cat} \rho_\te \left[ \rho_\tp \ex{-(\varepsilon+\mu)}- \rho_\ts\right]\} $. For the purpose of the analytical approach, we will assume that all the particles interact via hardcore repulsion. In this case, the virial coefficients simply read $\mathcal{B}_{ij} = \frac{4}{3}\pi \sigma_{ij}^3$, where $\sigma_{ij}$ denotes the distance of closest approach between particles of species $i$ and $j$.

We make several additional assumptions to simplify the analysis. First, we assume that substrate and product molecules only interact with enzymes, and not with one another, i.e. that $\sigma_{\ts\ts} =\sigma_{\tp\tp} =\sigma_{\ts\tp} =0$, which implies $\mathcal{B}_{\ts\ts} =\mathcal{B}_{\tp\tp} =\mathcal{B}_{\ts\tp} =0$. Second, as anticipated, substrate and product interact identically with the enzyme such that $\sigma_{\te\ts}=\sigma_{\te\tp}$, i.e. $\mathcal{B}_{\te\ts}=\mathcal{B}_{\te\tp}$. Third, we assume that the chemical reactions are fast relative to diffusion, so that we can employ a local equilibrium assumption (LEA). Within the LEA, the chemical equilibrium conditions ($R_\ts=R_\tp=0$) enslave the local densities of product and substrate to the local density of enzyme through $ \rho_{\ts}=\rho_{\ts\tp}\theta(\rho_{\te})$ and $ \rho_{\tp}=\rho_{\ts\tp}[1-\theta(\rho_{\te})]$,
where we have defined the total substrate and product density $\rho_{\ts\tp}\equiv \rho_\ts +\rho_\tp$ and the dimensionless partition factor
\begin{equation}
\theta(\rho_{\te})\equiv\frac{k_{\mathrm{spo}}+k_{\mathrm{cat}}\rho_{\te}e^{-(\varepsilon+\mu)}}{k_{\mathrm{spo}}+k_{\mathrm{cat}}\rho_{\te}e^{-(\varepsilon+\mu)}+k_{\mathrm{spo}}e^{\varepsilon}+k_{\mathrm{cat}}\rho_{\te}}.
\label{eq:theta}
\end{equation}
Summing the equations for $\rho_\ts$ and $\rho_\tp$ eliminates the reaction terms, and the total solute flux reads $-(D_\ts\nabla\rho_\ts+D_\tp\nabla\rho_\tp)-(D_\ts\rho_\ts+D_\tp\rho_\tp)\mathcal{B}_{\te\ts}\nabla\rho_\te$. Since $D_\ts$ and $D_\tp$ are constants, the first term is a total gradient, $-\nabla(D_\ts\rho_\ts+D_\tp\rho_\tp)$, and within the LEA $D_\ts\rho_\ts+D_\tp\rho_\tp=D_\ts^\text{eff}(\rho_\te)\rho_{\ts\tp}$, where $D_\ts^\text{eff}(\rho_\te)\equiv D_\ts[\theta(\rho_\te)+\alpha(1-\theta(\rho_\te))]$ is the population-averaged diffusion coefficient of a solute molecule. The LEA thus reduces the ternary mixture to an effective binary mixture of enzymes and solute molecules (substrate or product indistinguishably), obeying~\cite{SM}
\begin{align}
\partial_t\rho_\te &= \nabla\cdot \{D_\te\left[(1+\rho_\te\mathcal{B}_{\te\te})\nabla\rho_\te+\rho_\te\mathcal{B}_{\te\ts}\nabla\rho_{\ts\tp}\right]\},\label{eq:reduced_e}\\
\partial_t\rho_{\ts\tp} &= \nabla\cdot\left[\nabla\big(D_\ts^\text{eff}(\rho_\te)\rho_{\ts\tp}\big)+D_\ts^\text{eff}(\rho_\te)\rho_{\ts\tp}\mathcal{B}_{\te\ts}\nabla\rho_\te\right].\label{eq:reduced_sp}
\end{align}
Importantly, we note that the solute response to enzyme gradients includes, in addition to the excluded-volume contribution $\propto \mathcal{B}_{\te\ts}$, a contribution $\propto {D_\ts^\text{eff}}'(\rho_\te)$ (the prime denotes a derivative with respect to $\rho_\te$) due to the dependence of the effective diffusivity on enzyme density. 
Indeed, Eqs.~\eqref{eq:reduced_e}--\eqref{eq:reduced_sp} imply the effective diffusion matrix
\begin{equation}
    \boldsymbol{\mathcal{D}}^{\mathrm{LEA}}=\begin{pmatrix} D_\te (1+\rho_\te \mathcal{B}_{\te\te}) & D_\te\rho_\te \mathcal{B}_{\te\ts}\\ D_\ts^\text{eff}\rho_{\ts\tp}\mathcal{B}_{\ts\te}^\text{eff} & D_\ts^\text{eff}\end{pmatrix},
    \label{eq:Dlea2}
\end{equation}
with $\mathcal{B}^\text{eff}_{\ts\te}\equiv\mathcal{B}_{\te\ts}+\partial_{\rho_\te}\ln D_\ts^\text{eff}$, where the second term represents the additional contribution just discussed. An instability will occur when this matrix, evaluated at the homogeneous state $\rho_\te=\rho_\te^0$ and $\rho_{\ts\tp}=\rho_{\ts\tp}^0$,   loses its positive definiteness, i.e.~when $\det(\boldsymbol{\mathcal{D}}^{\mathrm{LEA}})<0$. At leading order in the densities and in the diameters of molecules, this yields the following instability criterion:
\begin{equation}
    X \frac{\sigma_{\te\ts}^3}{\sigma_{\te\te}^3}\frac{\alpha-1}{\theta+\alpha(1-\theta)} \frac{x(1-e^{-\mu})}{[1+e^\varepsilon+x(1+e^{-\varepsilon-\mu})]^2} >\frac{1}{8}+\phi_\te
    \label{eq:instab_crit}
\end{equation}
where we have introduced the dimensionless parameters  $x \equiv k_\text{cat}\rho^0_\te/k_\text{spo}$ (which measures the relative importance of the catalytic and spontaneous reaction pathways), $\phi_\te \equiv \rho_\te^0 v_e$ (volume fraction of enzyme, where $v_\text{e} \equiv \frac{4}{3}\pi (\frac{\sigma_{\te\te}}{2})^3$), $X \equiv \rho_{\ts\tp}^0 v_\text{e}$ (number of substrate and product molecules per volume of enzyme), and $\alpha \equiv D_\tp/D_\ts$ (diffusivity contrast). 

Eq.~\eqref{eq:instab_crit} is our main analytical result
 As the right hand side is positive, a necessary condition for the instability is that the left hand side be positive as well. CIPS thus requires $\mu(\alpha-1)>0$: it is impossible at equilibrium ($\mu=0$) or for equal diffusivities ($\alpha=1$), and, for $\mu>0$, the product must diffuse faster than the substrate ($\alpha>1$). Provided these necessary conditions are satisfied, CIPS can be achieved by sufficiently high $X$, i.e.~sufficiently high substrate and product concentration. Importantly, we predict that CIPS is possible even when the size of substrate and product is negligible relative to the size of the enzyme, so that they are effectively point-like (corresponding to the case $\sigma_{\te\ts} = \sigma_{\te\te}/2$), which distinguishes CIPS from phase separation induced by equilibrium depletion forces~\cite{Oosawa1954,Asakura1958,Vrij1976,Binder2014}.

The instability criterion (\ref{eq:instab_crit}) can be simplified in certain relevant limits. Consider the situation where both reaction pathways are strongly driven ($\mu+\varepsilon\to \infty$ and $\varepsilon\to-\infty$), 
and where the size of substrate and product molecules is negligible ($\sigma_{\te\ts} = \sigma_{\te\te}/2$). One finds that, for a fixed value of $\alpha$, there is a value of $x$ ($x_\text{max}=1/\sqrt{\alpha}$) that maximizes the left hand side of the inequality. At this optimal value of $x$, the instability criterion becomes $ X > (1+8\phi_\te)\frac{\sqrt{\alpha}+1}{\sqrt{\alpha}-1}$, which can only be fulfilled if $X>1+8\phi_\te$, even for $\alpha\to\infty$ (e.g.~when the substrate binds to a static matrix, $D_\ts\to0$). 

\emph{Comparison with the continuum theory.---} Eq.~\eqref{eq:instab_crit} can be compared directly with the instability criterion of Ref.~\cite{Cotton2022} (see Supplemental Material~\cite{SM} for details). In the solvent-free, incompressible limit considered there ($\phi_{\ts\tp}\to 1$ and $\phi_\te\ll1$), the two criteria become essentially identical, with the bare diffusion coefficients $D_\tp$ and $D_\ts$ playing the role of the off-diagonal coefficients $D_{\tp\te}$ and $D_{\ts\te}$. This correspondence has two consequences. First, 
the asymmetry postulated in Ref.~\cite{Cotton2022} is carried here by the self-diffusion coefficients alone, the enzymes responding identically to substrate and product. Second, it highlights the role of the solvent. In Ref.~\cite{Cotton2022}, the instability is enhanced by the size ratio $v_\te/v_\ts$, which is large provided the enzyme is larger than the solute. In a dilute solution, this factor is replaced by $X=\rho_{\ts\tp}^0 v_\te$, which is not large \textit{a priori}, and the substrate and product concentration becomes a control parameter in its own right. Note that, using $\rho_{\ts\tp}^0 v_s = \phi_{\ts\tp}$, $X$ can be rewritten as $X = \phi_{\ts\tp}v_\te/v_\ts$, highlighting the correspondence to Ref.~\cite{Cotton2022} in the limit $\phi_{\ts\tp}\to 1$. CIPS thus requires both sufficiently large diffusivity contrast and solute concentration (see above).
For an enzyme of radius $5~\text{nm}$, $X=1$ corresponds to a total substrate and product concentration of about $3~\text{mM}$, comparable to typical intracellular metabolite concentrations~\cite{Bennett2009,Park2016}. Finally, our microscopic approach allows us to quantify the competition between CIPS and depletion-induced interactions explicitly, as we now show with simulations.

\begin{figure*}
    \centering
    \includegraphics[width=0.66\columnwidth]{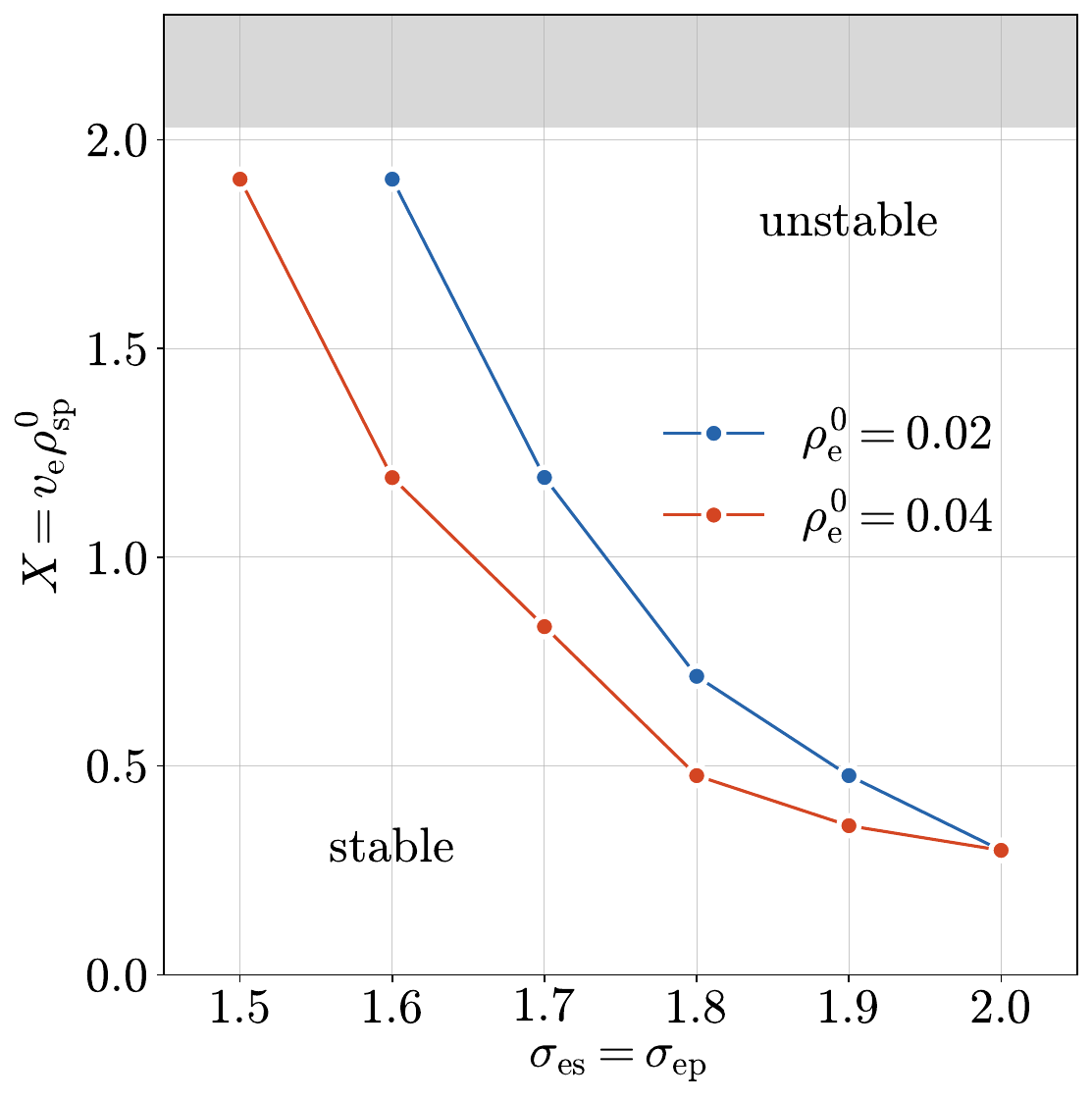}
    \includegraphics[width=0.66\columnwidth]{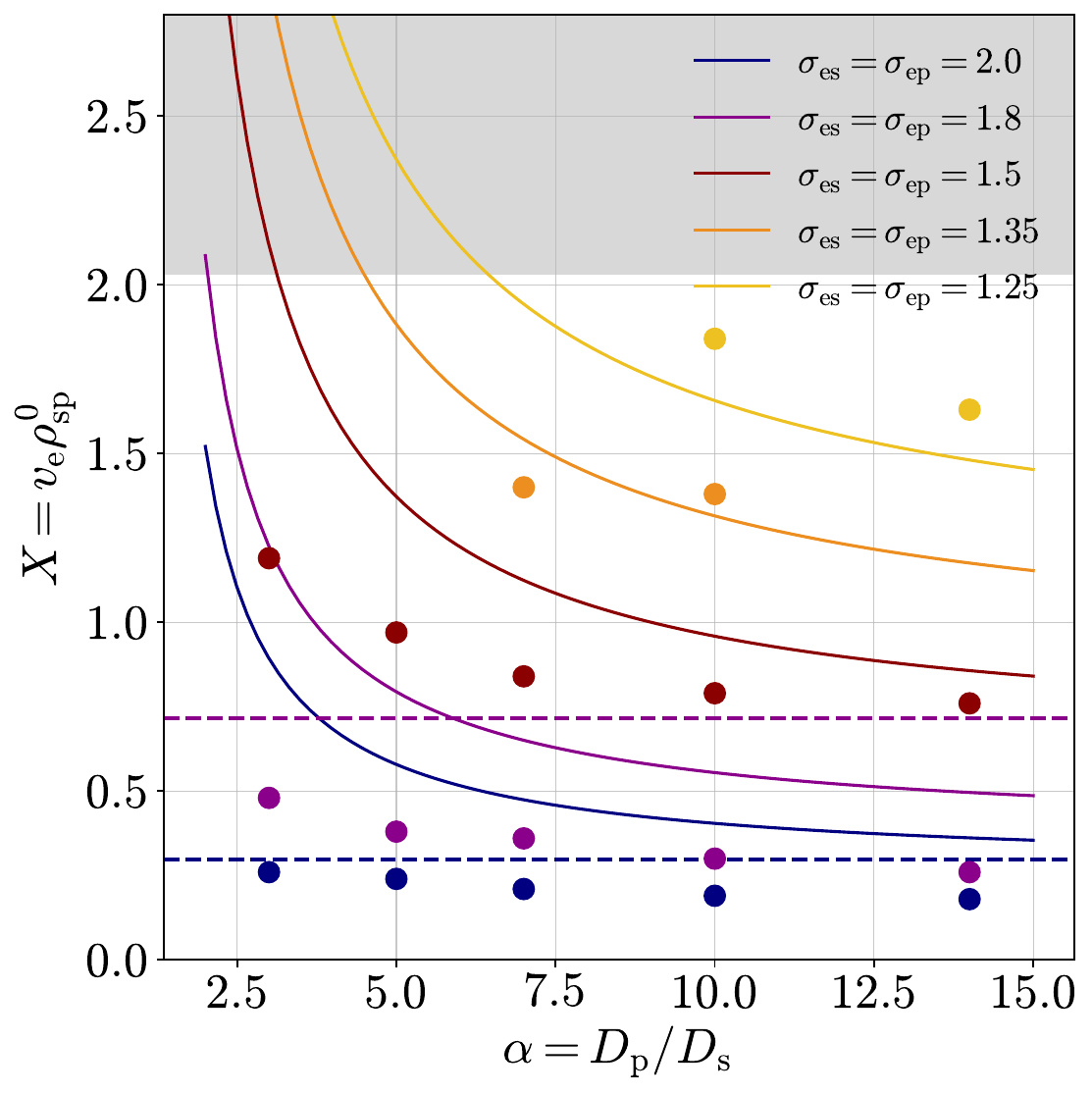}
    \includegraphics[width=0.66\columnwidth]
    {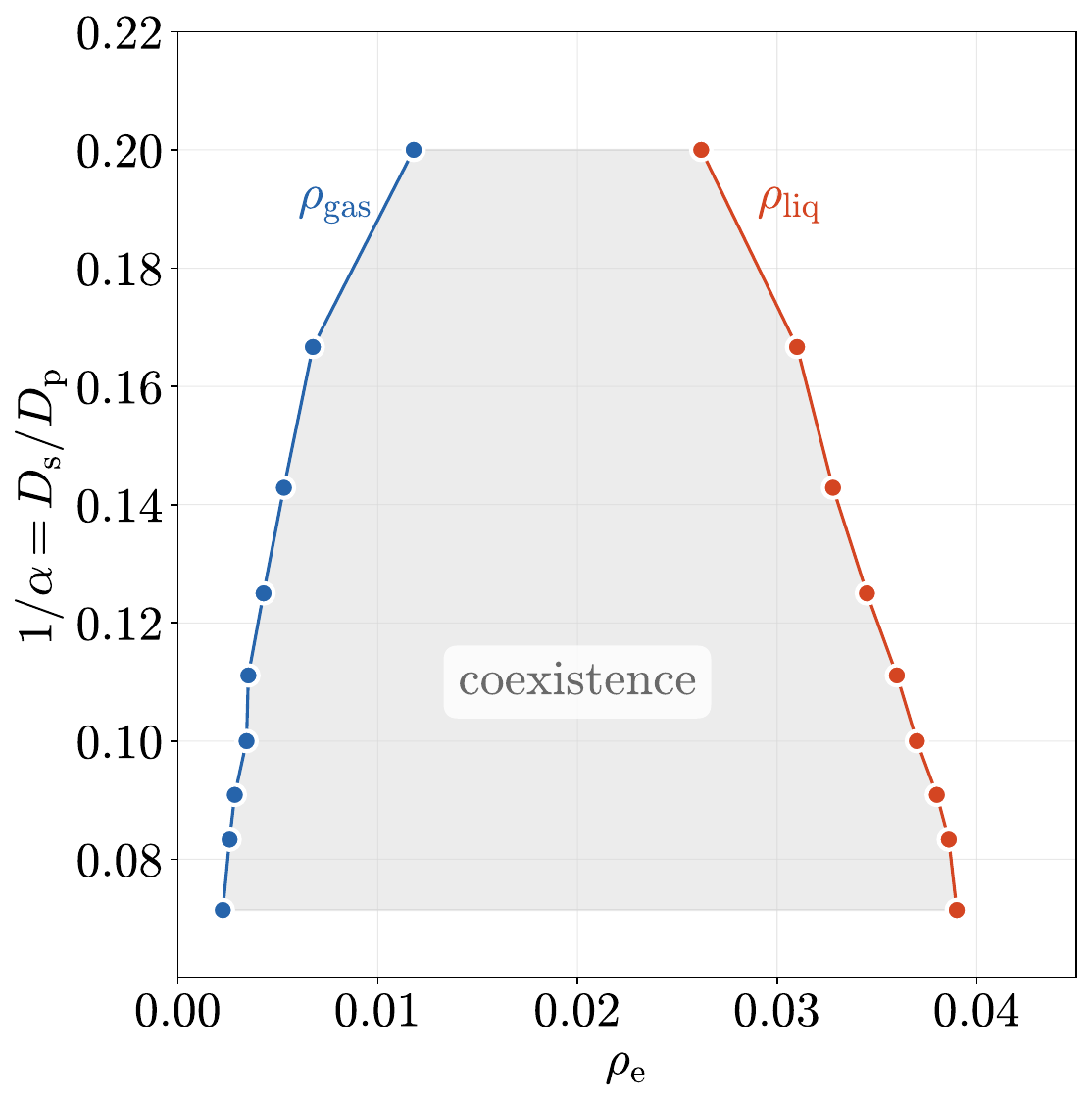}
    \caption{{Left: Stability diagram of the nonreactive mixture, for two values of $\rho_\te^0$, in the plane $(\sigma_{\te\ts},X)$. Middle: Stability diagram  with spontaneous and catalytic reactions, for $\rho_\te^0 =0.02$, in the plane $(\alpha,X)$, for different values of the interaction distance $\sigma_{\te\ts}$. Symbols: BD simulations; solid lines: instability criterion from Eq. \eqref{eq:instab_crit}; dashed lines:  depletion-induced instability of the nonreactive mixture [from (a)] for $\sigma_{\te\ts}=2.0$ and $1.8$. For smaller $\sigma_{\te\ts}$, depletion-induced instability occurs at values of $X$ not reached in simulations (grey domains). Right: Enzyme densities in the coexisting dense and dilute phases as a function of $1/\alpha=D_\ts/D_\tp$, at fixed $\sigma_{\te\ts}=1.5$ and $\rho_\te^0=0.02$.
    In all cases $\sigma_{\te\ts}=\sigma_{\te\tp}$, and remaining parameters are chosen as $\varepsilon=-8$, $\mu=16$, $D_\ts=D_\te$, $x=k_\text{cat}\rho_\te^0/k_\text{spo}=0.267$.} }
    \label{fig:BDresults}
\end{figure*}

\emph{Brownian dynamics simulations.---} We test these predictions with Brownian dynamics (BD) simulations of the reactions described above. BD simulations are more conveniently performed with smooth  rather than hardcore potentials, and we now assume that the particles interact via Weeks-Chandler-Andersen potentials (truncated and shifted Lennard-Jones)~\cite{Weeks1971}, which are such that $V_{ij}(r) = 4\varepsilon_\text{WCA} [ (\sigma_{ij}/r)^{12}-(\sigma_{ij}/r)^{6}] +\varepsilon_\text{WCA}$ if $r < 2^{1/6} \sigma_{ij}$ and 0 otherwise. 
We do not expect this choice to cause any qualitative difference with the predictions from our analytical calculation. We take the radius of the enzymes to be unity (i.e. $\sigma_{\te\te}=2$): this defines the unit length of the problem. We will consider two enzyme densities, namely $\rho_\text{e}^0=0.02$ and $0.04$, and the total density of substrate and product will vary from $\rho_{\ts\tp}^0=0.014$ to $0.48$. Finally, the $S \leftrightharpoons P $ interconversions are performed every $N_{\rm reac}$ steps of  the Euler-Maruyama integration of the equations of motion (see Supplemental Material~\cite{SM} for additional details on the numerical simulations and simulation snapshots).

Even without reactions, the mixture may be unstable, since depletion interactions, which cannot be accounted for in our low-density analytical approach, may come into play~\cite{Oosawa1954,Asakura1958,Vrij1976,Binder2014}. Such depletion interactions may be further enhanced in the present case, since we allow for the pair interactions to be nonadditive, i.e.~we allow $\sigma_{\te j}>(\sigma_{\te\te}+\sigma_{jj})/2$ for $j\in\{\text{s},\text{p}\}$~\cite{Louis2001,Roth2001,Roth2001a,Dijkstra1998}. In order to discriminate between depletion-induced phase separation and CIPS, we first plot on Fig.~\ref{fig:BDresults}(a) the phase diagram in the plane $( \sigma_{\te\ts}, X)$, for a non-reactive mixture with substrate only (the same diagram is obtained with reactions when $\mu=0$ or $\alpha=1$, see Supplemental Material~\cite{SM}). The critical value of $X$ above which the instability occurs is denoted by $X_\text{crit}$  and is estimated by calculating the Binder cumulant associated with density distributions~\cite{SM}.  Since $\sigma_{\ts\ts}=0$, the value  $\sigma_{\te\ts}=1$ corresponds to the case of an additive mixture with point-like solute, for which no depletion-induced phase separation occurs. For higher value of the cross diameter $\sigma_{\te\ts}$, we find that there exists a value of $X$ above which depletion-induced equilibrium phase separation may occur.

For reactive mixtures, we plot the phase diagrams for fixed values of  $\sigma_{\te\ts}=\sigma_{\te\tp}$  in the plane $(\alpha,X)$ on Fig.~\ref{fig:BDresults}(b)
(the time evolution of a homogeneous and of a phase-separating system is shown in Supplemental Movie 1 \cite{SM}, together with the distribution of local enzyme density).
Importantly, we observe that, when the diffusivity contrast $\alpha$ is large enough, phase separation occurs far outside of the region where equilibrium depletion-induced phase separation is expected to occur, which is above the dashed lines for $\sigma_{\te\ts}=2.0,1.8$ or within the grey region for smaller values of $\sigma_{\te\ts}$.
We also plot  the critical value of $X$ above which CIPS is predicted to occur from Eq. \eqref{eq:instab_crit}, and we find a good semi-quantitative, parameter-free comparison to the simulation results.

For a particular set of values ($\rho_\te^0=0.02$, $x=0.267$, $\sigma_{\te\ts}=1.5$ and $X=1.19$), we measured the density of the coexisting enzyme-poor and enzyme-rich phases ($\rho_\text{gas}$ and $\rho_\text{liq}$, respectively) as a function of $\alpha$ [Fig.~\ref{fig:BDresults}(c)]. They display an analogous form to an equilibrium liquid-gas coexistence diagram, the difference being that, in the nonequilibrium situation we consider here, the control parameter is not the strength of attractive interactions but the diffusivity contrast $\alpha = D_\tp/D_\ts$.

\emph{Mechanism and nonreciprocal couplings.---} Eqs.~\eqref{eq:reduced_e}--\eqref{eq:reduced_sp} make the mechanism of CIPS explicit. Enzymes respond identically to substrate and product, and are pushed down gradients of the total solute density by excluded-volume forces only. The nonequilibrium character of the dynamics is entirely contained in the solute equation, through a flux of the form $-\nabla[D_\ts^\text{eff}(\rho_\te)\rho_{\ts\tp}]$, generated by nonequilibrium enzyme catalysis, which makes the effective solute diffusivity depend on the local enzyme density and causes the solute to accumulate in regions of low effective diffusivity, as in models of motility-regulated (or quorum-sensing) active particles, which undergo motility-induced phase separation~\cite{Tailleur2008,Cates2015}. Indeed, whether $D_\ts^\text{eff}(\rho_\te)$ increases or decreases with increasing enzyme density depends on the sign of $\mu(\alpha-1)$.
For $\mu(\alpha-1)>0$, solute molecules experience faster effective diffusivity in the presence of the enzyme, and therefore accumulate in enzyme-poor regions. The resulting excess of solute in enzyme-poor regions pushes enzymes toward enzyme-rich regions, which closes a positive feedback loop [Fig.~\ref{fig:system_sketch}(c)]. This mechanism is reminiscent of the self-attraction of chemically active, chemotactic particles~\cite{AgudoCanalejo2019}, the enzymes acting here as effective sinks of solute, and their chemotactic response being due to excluded volume only.

The off-diagonal elements of $\boldsymbol{\mathcal{D}}^{\mathrm{LEA}}$ [Eq.~\eqref{eq:Dlea2}] are $D_\te\rho_\te\mathcal{B}_{\te\ts}$ (response of the enzymes to solute gradients) and $D_\ts^\text{eff}\rho_{\ts\tp}\mathcal{B}^\text{eff}_{\ts\te}$ (response of the solute to enzyme gradients) where, explicitly,
\begin{equation}
	\mathcal{B}_{\ts\te}^\text{eff} = \mathcal{B}_{\te\ts} + \frac{k_\mathrm{cat}}{k_\mathrm{spo}}\frac{\alpha-1}{\theta+\alpha(1-\theta)} \frac{1-e^{-\mu}}{[1+e^\varepsilon+x(1+e^{-\varepsilon-\mu})]^2}.
\end{equation}
Provided that $\mu(\alpha-1) \neq 0$, i.e.~that there is both a nonequilibrium drive and a diffusivity contrast, solute molecules then respond to enzymes differently from how enzymes respond to solute molecules: the effective enzyme--solute couplings are nonreciprocal~\cite{Saha2020,You2020,Duan2023,Dinelli2023,Fruchart2026}. 

This picture can be reconciled with the effective free energy derived in Ref.~\cite{Cotton2022}. Indeed, setting both fluxes in Eqs.~\eqref{eq:reduced_e}--\eqref{eq:reduced_sp} to zero enslaves the solute density to the enzyme density, $\rho_{\ts\tp}(\rho_\te)\propto e^{-\mathcal{B}_{\te\ts}\rho_\te}/D_\ts^\text{eff}(\rho_\te)$, up to a prefactor fixed by solute conservation. The enzyme density then obeys $\nabla\mu_\text{eff}=0$, with the effective chemical potential 
$\beta \mu_\text{eff}(\rho_\te)=\ln\rho_\te+\mathcal{B}_{\te\te}\rho_\te+\mathcal{B}_{\te\ts}\rho_{\ts\tp}(\rho_\te)$, which is the dilute, solvent-containing counterpart of Eq.~(8) in Ref.~\cite{Cotton2022}. The spinodal condition $\mu_\text{eff}'(\rho_\te)<0$ reads $1+\rho_\te\mathcal{B}_{\te\te}<\rho_\te\rho_{\ts\tp}\mathcal{B}_{\te\ts}\mathcal{B}^\text{eff}_{\ts\te}$: it coincides with $\det\boldsymbol{\mathcal{D}}^\text{LEA}<0$, and reduces to Eq.~\eqref{eq:instab_crit} when the depletion term $\rho_\te\rho_{\ts\tp}\mathcal{B}_{\te\ts}^2$ is neglected. Thus, as long as the off-diagonal elements of $\boldsymbol{\mathcal{D}}^\text{LEA}$ have the same sign, which is the case in the CIPS regime ($\mu(\alpha-1)>0$), the onset of the instability is equilibrium-like. In this regime, nonreciprocity affects the dynamics (relaxation, coarsening, fluctuations) rather than the instability threshold.

Nonreciprocity is expected to play a more prominent role when the off-diagonal elements of $\boldsymbol{\mathcal{D}}^\text{LEA}$ have opposite signs, which occurs when the diffusivity contrast is reversed ($\mu(\alpha-1)<0$).
Then $D_\ts^\text{eff}(\rho_\te)$ decreases with enzyme density, such that $\mathcal{B}^\text{eff}_{\ts\te}$ can become negative, and solute molecules accumulate in enzyme-rich regions while enzymes are repelled by them, a chase-and-run interaction~\cite{Saha2020,You2020}. The spinodal condition above then predicts that catalysis opposes, and may eventually suppress, depletion-induced phase separation, i.e.~that sufficiently strong catalytic activity could dissolve enzyme condensates held together by equilibrium interactions. This opens rich perspectives for molecular simulations of biomolecular condensates, which have recently attracted much attention~\cite{Dignon2020,Joseph2021,Maristany2025}.

\bibliography{refs.bib}

\end{document}


\setcounter{secnumdepth}{3}

\beginsupplement

\title{Catalysis-induced phase separation of enzymes \emph{via} diffusivity contrast \\ \textit{Supplemental Material}}

\author{Marie Jardat}
\affiliation{Sorbonne Universit\'e, CNRS, Physical Chemistry of Electrolytes and Interfacial Nanosystems (PHENIX), 4 place Jussieu, Paris, France}

\author{Jacques Fries}
\affiliation{Sorbonne Universit\'e, CNRS, Physical Chemistry of Electrolytes and Interfacial Nanosystems (PHENIX), 4 place Jussieu, Paris, France}

\author{Vincent Dahirel}
\affiliation{Sorbonne Universit\'e, CNRS, Physical Chemistry of Electrolytes and Interfacial Nanosystems (PHENIX), 4 place Jussieu, Paris, France}

\author{Jaime Agudo-Canalejo}
\email{j.agudo-canalejo@ucl.ac.uk}
\affiliation{Department of Physics and Astronomy, University College London, London WC1E 6BT, United Kingdom}

\author{Pierre Illien}
\email{pierre.illien@sorbonne-universite.fr}
\affiliation{Sorbonne Universit\'e, CNRS, Physical Chemistry of Electrolytes and Interfacial Nanosystems (PHENIX), 4 place Jussieu, Paris, France}

\date{\today}

\maketitle

\onecolumngrid

\tableofcontents

\section{Analytical approach in the low-density limit}

\subsection{Low density approximation of the Smoluchowski equation}

We assume here that there are $N_i$ particles of each species, and we denote by $\rr^i_1,\dots,\rr^i_{N_i}$ the coordinates of the particles of species $i$ ($i\in \{E,S,P\}$). The total number of particles is $N=N_\te+N_\ts+N_\tp$. We denote by $V_{ij}(\rr^i_\alpha-\rr^j_\beta)$ the interaction potential of two particles of species $i$ and $j$ of respective positions $\rr^i_\alpha$ and $\rr^j_\beta$ ($\alpha\in\{1,\dots,N_i\}$ and $\beta\in\{1,\dots,N_j\}$). In the absence of chemical reactions, and denoting by $\rr^N$ the $3N$-dimensional vector that contains the position of all particles of the system, the probability $P$ to observe a given configuration $\rr^N$ evolves as \cite{McQuarrie1976,Dhont1996}
\begin{equation}
	\frac{\partial}{\partial t} P(\rr^N;t) = \Omega(\rr^N) P(\rr^N;t) ,
	\label{Smolu}
\end{equation}
where the Smoluchowski operator $\Omega$ is defined as
\begin{equation}
\Omega(\rr^N) =	\sum_i \sum_{\alpha=1}^{N_i} \left[\frac{\partial}{\partial \rr^i_\alpha} - \frac{\partial}{\partial \rr^i_\alpha} \beta U(\rr^N)\right]D_i \frac{\partial}{\partial \rr^i_\alpha},
\end{equation}
where $\beta = (\kB T)^{-1}$.  The potential energy $U$ includes all pairwise interactions:
\begin{equation}
	U(\rr^N) = \frac{1}{2}\sum_{i\in\{\te,\ts,\tp\}} \sum_{\alpha\neq \beta} ^{N_i} V_{ii} (\rr^i_\alpha- \rr^i_\beta)
	+ \sum_{i\in\{\te,\ts,\tp\}}\sum_{j\neq i} \sum_{\alpha =1}^{N_i}\sum_{\beta =1}^{N_j} V_{ij}(\rr_\alpha^i - \rr_\beta^j).
\end{equation}
Let $p_1^i(\rr,t)$ be the probability to find a particle of species $i$ at time $t$ (the number density will be deduced as $\rho_i(\rr,t) = N_i p_1^i(\rr,t)$). Integrating Eq. \eqref{Smolu} over the positions of all particles except one of species $i$ yields (see e.g. \cite{Grosberg2015} for a similar procedure), one gets
\begin{eqnarray}
	\frac{\partial}{\partial t}p_1^i(\rr,t) 
	&=& D_i \nabla_{\rr}^2 p_1^i(\rr,t)
	+\frac{D_i}{\kB T} (N_i-1) \nabla_{\rr} \int_{\mathcal{V}} \dd \rr'\; p_2^{ii}(\rr,\rr',t) \nabla_{\rr}V_{ii}(\rr-\rr')\nonumber\\
	&&+\frac{D_i}{\kB T}\sum_{j\neq i}  N_j \nabla_{\rr}\int_{\mathcal{V}} \dd \rr'\; p_2^{ij}(\rr,\rr',t) \nabla_{\rr}V_{ij}(\rr-\rr')
	\label{eq_p1}
\end{eqnarray}
where $p_2^{ij}(\rr,\rr';t)$ is the joint probability to find a particle of species $i$ (resp. $j$) at position $\rr$ (resp. $\rr'$) at time $t$, and where $\mathcal{V}$ is the volume of the system. The evolution equation of this quantity can formally be written by integrating the Smoluchowski equation [Eq. \eqref{Smolu}] over all positions but two. However, this equation will involve three-particle probabilities, and so on. 

In the low-density limit, this hierarchy of equations can be closed as follows. The two- and one-particle distributions are related through the relation \cite{McQuarrie1976}:
\begin{equation}
	p_2^{ij} (\rr,\rr';t) = p_1^{i} (\rr;t)p_1^{j} (\rr';t)g^{ij}(\rr-\rr')
\end{equation}
where $g^{ij}(\rr-\rr')$ is the pair distribution function. In the low-density limit, and at linear order in the densities $\rho_i^0$, the pair distribution functions are known exactly and read:
\begin{equation}
	g^{ij}(\rr-\rr') = \ex{-\beta V_{ij}(\rr-\rr')}.
\end{equation}
Eq. \eqref{eq_p1} then becomes
\begin{eqnarray}
	\frac{\partial}{\partial t}p_1^i(\rr,t) 
	&=& D_i \nabla_{\rr}^2 p_1^i(\rr,t)
	+\frac{D_i}{\kB T} (N_i-1) \nabla_{\rr}\left[ p_1^i(\rr;t)\int_{\mathcal{V}} \dd \rr'\; p_1^{i}(\rr',t) \underbrace{\ex{-\beta V_{ii}(\rr-\rr')} \nabla_{\rr}V_{ii}(\rr-\rr')}_{=\kB T \nabla_{\rr'} \ex{-\beta V_{ii}(\rr-\rr')}} \right]\nonumber\\
	&&+\frac{D_i}{\kB T}\sum_{j\neq i}  N_j \nabla_{\rr}\left[p_1^{i}(\rr,t)\int_{\mathcal{V}} \dd \rr'\; p_1^{j}(\rr',t)\underbrace{\ex{-\beta V_{ij}(\rr-\rr')} \nabla_{\rr}V_{ij}(\rr-\rr')}_{=\kB T \nabla_{\rr'} \ex{-\beta V_{ij}(\rr-\rr')}}\right].
\end{eqnarray}
Integrating by parts yields
\begin{eqnarray}
	\frac{\partial}{\partial t}p_1^i(\rr,t) 
	&=& D_i \nabla_{\rr}^2 p_1^i(\rr,t)
	+D_i (N_i-1) \nabla_{\rr}\left[ p_1^i(\rr;t)\int_{\mathcal{V}} \dd \rr'\;(1-\ex{-\beta V_{ii}(\rr-\rr')}) \nabla_{\rr'} p_1^{i}(\rr',t) \right]\nonumber\\
	&&+D_i\sum_{j\neq i}  N_j \nabla_{\rr}\left[p_1^{i}(\rr,t)\int_{\mathcal{V}} \dd \rr'\; (1-\ex{-\beta V_{ii}(\rr-\rr')}) \nabla_{\rr'}p_1^{j}(\rr',t)\right],
\end{eqnarray}
where we used the Einstein relations $D_i = \mu_i \kB T$. Since the integrands in the integrals over $\rr'$ are short-ranged, we can expand $p_1^i(\rr',t)$ and $p_1^j(\rr',t)$ around $\rr=\rr'$ \cite{Felderhof1978}. Finally, in the thermodynamic limit where $N_i\gg 1$ and $\mathcal{V}\to\infty$ with fixed densities $\rho_i^0=N_i/\mathcal{V}$, and recalling that $\rho_i(\rr,t) = N_i p_i(\rr,t)$, we get:
\begin{eqnarray}
	\frac{\partial}{\partial t}\rho_i(\rr,t) 
	&=& D_i \nabla_{\rr}^2 \rho_i(\rr,t)
	+D_i  \nabla_{\rr}\left[ \rho_i(\rr;t) \sum_j \mathcal{B}_{ij}\nabla_{\rr} \rho_j(\rr,t) \right],
\end{eqnarray}
where we defined the virial coefficient:
\begin{equation}
	\mathcal{B}_{ij} = \int \dd \rr [1-\ex{-\beta V_{ij}(\rr)}].
\end{equation}
If we further assume that the particles interact \emph{via} hardcore interactions, the virial coefficients are simply given by
\begin{equation}
	\mathcal{B}_{ij} = \frac{4}{3} \pi \sigma_{ij}^3. 
\end{equation}
Adding the effect of chemical reactions then yields the reaction-diffusion equations:
\begin{equation}
    \partial_t \rho_i = \nabla \cdot \sum_{j}\mathcal{D}_{ij} \nabla \rho_j + R_i
\end{equation}
where the $R_i$ are the reaction fluxes, recalled from the main text for completeness:
\begin{eqnarray}
    R_\te&=&0 \\
    R_{\ts,\tp} &=& \pm\{	 k_\text{spo} \left[ \rho_\tp-\ex{\varepsilon} \rho_\ts \right]
+ k_\text{cat} \rho_\te \left[ \rho_\tp \ex{-(\varepsilon+\mu)}- \rho_\ts\right]\}  
\end{eqnarray}
and where the diffusion matrix is
\begin{equation}
	\boldsymbol{\mathcal{D}} =
	\begin{pmatrix}
		D_\te(1+\rho_\te\mathcal{B}_{\te\te}) & D_\te \rho_\te\mathcal{B}_{\te\ts}  & D_\te \rho_\te\mathcal{B}_{\te\tp}  \\
		D_\ts \rho_\ts \mathcal{B}_{\te\ts} & D_\ts(1+\rho_\ts\mathcal{B}_{\ts\ts}) & D_\ts \rho_\ts \mathcal{B}_{\ts\tp} \\
		D_\tp \rho_\tp \mathcal{B}_{\te\tp} & D_\tp \rho_\tp \mathcal{B}_{\ts\tp} & D_\tp(1+\rho_\tp\mathcal{B}_{\tp\tp})
	\end{pmatrix}.
\end{equation}
This is the expression given in the main text.

\subsection{Details on local equilibrium approximation}

Assuming $\mathcal{B}_{\ts\ts}=\mathcal{B}_{\tp\tp}=\mathcal{B}_{\ts\tp}=0$, which follows from the assumption $\sigma_{\ts\ts}=\sigma_{\tp\tp}=\sigma_{\ts\tp}=0$, the evolution equations for the conserved quantities $\rho_\te$ and $\rho_{\ts\tp} = \rho_\ts+\rho_\tp$ read
\begin{equation}
	\begin{aligned}
		\partial_{t}\rho_{\te} &= \nabla\cdot\left[D_{\te}(1+\rho_\te^0 \mathcal{B}_{\te\te})\nabla\rho_{\te}+D_{\te}\rho_\te^0 \mathcal{B}_{\te\ts}\nabla\rho_{\ts}+D_{\te}\rho_\te^0 \mathcal{B}_{\te\tp}\nabla\rho_{\tp}\right],\\
		\partial_{t}\rho_{\ts\tp} &= \nabla\cdot\left[ (D_\ts \rho_\ts^0 \mathcal{B} _{\te\ts} + D_\tp \rho_\tp^0 \mathcal{B} _{\te\tp})\nabla\rho_{\te}+D_{\ts}\nabla\rho_{\ts}+D_{\tp}\nabla\rho_{\tp}\right].
	\end{aligned}
	\label{eq:conserved_quantities}
\end{equation}

Within the local equilibrium approximation (LEA), we get:
\begin{equation}
	\rho_{\ts}\simeq\rho_{\ts\tp}\theta(\rho_{\te})\ \ \text{and}\ \ \rho_{p}\simeq\rho_{\ts\tp}[1-\theta(\rho_{\te})],
	\label{eq:LEA}
\end{equation}
where we recall the expression of the partition factor $\theta(\rho_\te)$ for completeness:
\begin{equation}
\theta(\rho_{\te})\equiv\frac{k_{\mathrm{spo}}+k_{\mathrm{cat}}\rho_{\te}e^{-(\varepsilon+\mu)}}{k_{\mathrm{spo}}+k_{\mathrm{cat}}\rho_{\te}e^{-(\varepsilon+\mu)}+k_{\mathrm{spo}}e^{\varepsilon}+k_{\mathrm{cat}}\rho_{\te}}.
\end{equation}
One deduces
\begin{equation}
\nabla\rho_{\ts}\simeq\theta\nabla\rho_{\ts\tp}+\rho_{\ts\tp}\theta^{\prime}\nabla\rho_{\te}\ \ \text{and}\ \ \nabla\rho_{p}\simeq(1-\theta)\nabla\rho_{\ts\tp}-\rho_{\ts\tp}\theta^{\prime}\nabla\rho_{\te}.
	\label{eq:gradients_LEA}
\end{equation}
Introducing these into (\ref{eq:conserved_quantities}), and using $\mathcal{B}_{\te\ts}=\mathcal{B}_{\te\tp}$, which follows from $\sigma_{\te\ts}=\sigma_{\te\tp}$, we can express the evolution of the conserved quantities as $\partial_t \rho_i = \nabla \cdot \sum_j \mathcal{D}_{ij}^\text{LEA} \nabla \rho_j$, with the effective diffusion matrix
\begin{equation}
    \boldsymbol{\mathcal{D}}^\text{LEA} = 
    \begin{pmatrix}
    D_\te(1+\rho_\te \mathcal{B}_{\te\te}) & D_\te \rho_\te \mathcal{B}_{\te\ts} \\
    \rho_{\ts\tp} \mathcal{B}_{\te\ts}[D_\ts\theta+D_\tp(1-\theta)]+(D_\ts-D_\tp)\rho_{\ts\tp}\theta'
    & D_\ts\theta+D_\tp(1-\theta)
    \end{pmatrix} 
\end{equation}
which is also given in the main text, in a slightly different form. The instability condition is given by $\text{det}(\boldsymbol{\mathcal{D}}^\text{LEA})<0$ evaluated in the uniform state $\rho_\te=\rho_\te^0$ and $\rho_{\ts\tp}=\rho_{\ts\tp}^0$. We simplify the determinant of $\boldsymbol{\mathcal{D}}^\text{LEA}$ by assuming the low-density limit, self-consistently with the derivation of the evolution equations, so that terms of the form $\rho_i\mathcal{B}_{jk}$ are considered to be of $\mathcal{O}(\epsilon)$, while $\theta$ and $\rho_\te \theta'$ are both $\mathcal{O}(1)$. In particular, neglecting one term of $\mathcal{O}(\rho_{\ts\tp}\rho_\te \mathcal{B}_{\te\ts}^2)=O(\epsilon^2)$, we obtain the instability criterion given in the main text.

\subsection{Comparison with field-based description of CIPS}

In this Section, we compare the results from the present approach to that obtained by Cotton \textit{et al.} \cite{Cotton2022} through a field-based description. We note that the no-solvent instability (Eq. 7) in Ref. \cite{Cotton2022} can be, by substituting the expressions for $R_{p}$ and $R_{\ts}$ and comparing with Eq. 6 in that reference, rewritten as
\begin{equation}
	\frac{1}{v_{\ts}}\frac{D_{\tp\te}-D_{\ts\te}}{D_{\ts\te}\theta+D_{\tp\te}(1-\theta)}\frac{k_{\mathrm{spo}}\hat{{k}}_{\mathrm{cat}}(1-e^{-\mu})}{(k_{\mathrm{spo}}+\hat{k}_{\mathrm{cat}}\phi_{\te}e^{-(\varepsilon+\mu)}+k_{\mathrm{spo}}e^{\varepsilon}+\hat{k}_{\mathrm{cat}}\phi_{\te})^{2}}>\frac{1}{v_{\te}\phi_{\te}}+\frac{1}{v_{\ts}(1-\phi_{\te})}
	\label{eq:cotton_instability}
\end{equation}
with
\begin{equation}
	\theta=\frac{k_{\mathrm{spo}}+\hat{{k}}_{\mathrm{cat}}\phi_{\te}e^{-(\varepsilon+\mu)}}{k_{\mathrm{spo}}+\hat{k}_{\mathrm{cat}}\phi_{\te}e^{-(\varepsilon+\mu)}+k_{\mathrm{spo}}e^{\varepsilon}+\hat{k}_{\mathrm{cat}}\phi_{\te}}
	\label{eq:cotton_theta}
\end{equation}
Multiplying both sides by $v_{\te}\phi_{\te}$, noting that $\hat{{k}}_{\mathrm{cat}}\phi_{\te}=k_{\mathrm{cat}}\rho_{\te}$, and taking the limit of small enzyme volume fraction, $\phi_{\te}\ll 1$, we get
\begin{equation}
	\frac{v_{\te}}{v_{\ts}}\frac{D_{\tp\te}-D_{\ts\te}}{D_{\ts\te}\theta+D_{\tp\te}(1-\theta)}\frac{k_{\mathrm{spo}}k_{\mathrm{cat}}\rho_{\te}(1-e^{-\mu})}{(k_{\mathrm{spo}}+k_{\mathrm{cat}}\rho_{\te}e^{-(\varepsilon+\mu)}+k_{\mathrm{spo}}e^{\varepsilon}+k_{\mathrm{cat}}\rho_{\te})^{2}}>1
	\label{eq:cotton_limit}
\end{equation}
This can be compared with Eq.~5 in the main text, which for the additive case $\sigma_{\te\ts}=\sigma_{\te\te}/2$ can be rewritten as
\begin{equation}
	\frac{v_{\te}}{v_{\ts}}\phi_{\ts\tp}\frac{D_{\tp}-D_{\ts}}{D_{\ts}\theta+D_{\tp}(1-\theta)}\frac{k_{\mathrm{spo}}k_{\mathrm{cat}}\rho_{\te}(1-e^{-\mu})}{(k_{\mathrm{spo}}+k_{\mathrm{cat}}\rho_{\te}e^{-(\varepsilon+\mu)}+k_{\mathrm{spo}}e^{\varepsilon}+k_{\mathrm{cat}}\rho_{\te})^{2}}>1+8\phi_\te
	\label{eq:our_instability_rewritten}
\end{equation}
where $\phi_{\ts\tp}=\rho_{\ts\tp}v_\ts$ is the combined substrate and product volume fraction.
Note that Eq.~\eqref{eq:cotton_limit} is derived for an incompressible mixture of e+s+p without solvent and low enzyme volume fraction, and thus corresponds to the limit $\phi_{\te}\ll 1$ and $\phi_{\ts\tp}\rightarrow 1$. In this limit, (\ref{eq:our_instability_rewritten}) becomes
\begin{equation}
	\frac{v_{\te}}{v_{\ts}}\frac{D_{\tp}-D_{\ts}}{D_{\ts}\theta+D_{\tp}(1-\theta)}\frac{k_{\mathrm{spo}}k_{\mathrm{cat}}\rho_{\te}(1-e^{-\mu})}{(k_{\mathrm{spo}}+k_{\mathrm{cat}}\rho_{\te}e^{-(\varepsilon+\mu)}+k_{\mathrm{spo}}e^{\varepsilon}+k_{\mathrm{cat}}\rho_{\te})^{2}}>1
	\label{eq:our_instability_rewritten_limit}
\end{equation}
The two expressions \eqref{eq:cotton_limit} and \eqref{eq:our_instability_rewritten_limit} become essentially identical in this limit, with $D_\tp$ and $D_\ts$ playing the role of $D_{\tp\te}$ and $D_{\ts\te}$, respectively.

\section{Numerical simulations}
\subsection{Particle-based simulations}
Particle-based simulations were carried out using the LAMMPS computational package~\cite{Thompson2022}, using the version released on 2 Aug 2023 - Update 2, with the default \texttt{lj} units. 
We use the command \texttt{fix brownian} of the LAMMPS package, that integrates the overdamped Langevin equations for the positions $\{\rr_i\}$ of particles using a forward Euler-Maruyama scheme~\cite{Allen1987}:
\begin{equation}
   \rr_i(t+\delta t) =\rr_i(t) + \sqrt{2D_i \delta t}  \boldsymbol{\zeta} -\delta t\frac{D_i}{\kB T} \sum_{j\neq i}\nabla V_{ij}  (r_{ij}).
    \label{overdampedLangevin_disc}
\end{equation}
where  $\boldsymbol{\zeta}$ is a random vector drawn from a normal distribution of mean $0$ and unit variance. 
The energy parameter of the WCA interaction potential is $\varepsilon_{\text{WCA}}=1$ where the energy unit is $k_{\text B}T$. The unit length is $\sigma_{\te\te}/2$, and we take $D_\te=0.5$, so that the time unit is $\tau_0=(\sigma_{\text ee}/2)^2/(2D_{\text e})$.

To ensure that substrate and product molecules only interact with enzymes, and not with one another, i.e. that $\sigma_{\ts\ts} =\sigma_{\tp\tp} =\sigma_{\ts\tp} =0$, we use the command \texttt{pair\_style zero}. $S$ and $P$ particles may thus be very close to each other without interacting.
However, since $S$ and $P$ both interact with $E$ via a WCA potential, increasing the interaction range $\sigma_{\ts\te} = \sigma_{\tp\te}$ (varied from $1.25$ to $2.0$ in our study) can lead to numerical instabilities. When multiple $S$ or $P$ particles are in close proximity to an $E$ particle, the repulsive forces add up, resulting in large net forces on $E$. This causes LAMMPS to report \texttt{lost atoms} errors if the integration time step $\delta t$ is too large. To mitigate this issue, we had to reduce $\delta t$ significantly as $\sigma_{\ts\te}$ or the density of  substrate/products increases. As a consequence, the accessible density range for  substrate/products is limited: $X=v_\te\rho_{\rm sp}^0$ is in any case lower than $2.05$. Large densities would require very small time steps, and thus a large computation time, to ensure numerical stability.
The time step $\delta t$ ranges from $1.0 \times 10^{-5}$ to $2.0 \times 10^{-4}$, depending on the system.
The simulations were performed using a cubic box of fixed length, $L_{\rm box} = 32.76$, with periodic boundary conditions, by varying the number of $E$ particles to adjust $\rho_{\te}^0$, and the initial number of $S$ and $P$ particles to adjust $\rho_{\rm sp}^0$. The total duration of each trajectory is $3\times10^4$. 
The $S \leftrightharpoons P$ interconversions are performed every $N_{\rm reac}$ time steps of the Euler-Maruyama integration of the equations of motion, with $\tau_{\rm reac} = N_{\rm reac} \delta t = 2.0 \times 10^{-2}$ in all cases. We have taken $k_{\rm spo}=0.05$ in all cases.

To implement the interconversions, we use PyLammps, the wrapper Python class for LAMMPS. With the
command \texttt{find pair neighlist}, for each reacting particle ($S$ or $P$), we get the number 
of $E$ particles located at a distance smaller than $r_{\rm local}=\sigma_{\te\te}/2+\sigma_{\te\tp}$, and compute the density $\rho_\te$ of $E$ particles in the sphere of radius $r_{\rm local}$ around it.  At each reaction attempt, for each $S$ particle, we draw a uniform random number $g \in [0,1]$ and compare it to the reaction probability
\begin{equation}
P_{S \to P} = \tau_{\rm reaction} \left(k_{\rm spo} e^\varepsilon + k_{\rm cat} \rho_\te\right),
\end{equation}
Similarly, for each $P$ particle, the reverse reaction $P \to S$ occurs with probability
\begin{equation}
P_{P \to S} = \tau_{\rm reaction}\left(k_{\rm spo} + k_{\rm cat} \rho_\te e^{-(\varepsilon+ \mu)}\right),
\end{equation}
If the random number $g$ is smaller than the respective probability, the particle type is updated, and the Brownian dynamics parameters are reset with new random seeds to avoid correlations.
The updated particle types are then reassigned to their respective LAMMPS groups defining $S$ and $P$ particles, before resuming the simulation. The populations of $S$ and $P$ particles converge to a stationary state at very short time (after $500\,\tau_0$ in any case).

\begin{figure}
    \centering
    \includegraphics[width=0.45\linewidth]{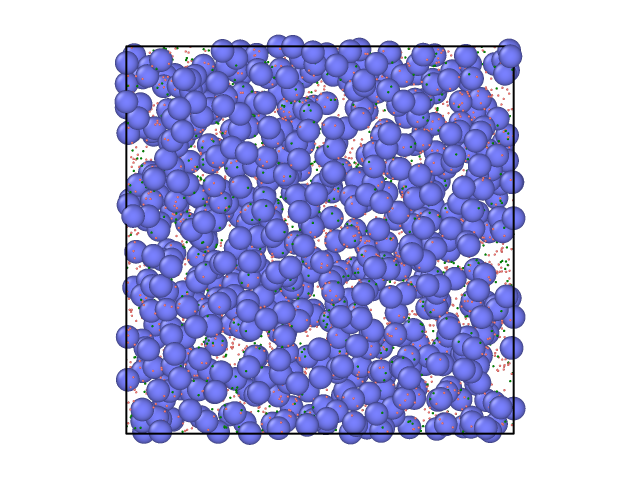}
    \includegraphics[width=0.45\linewidth]{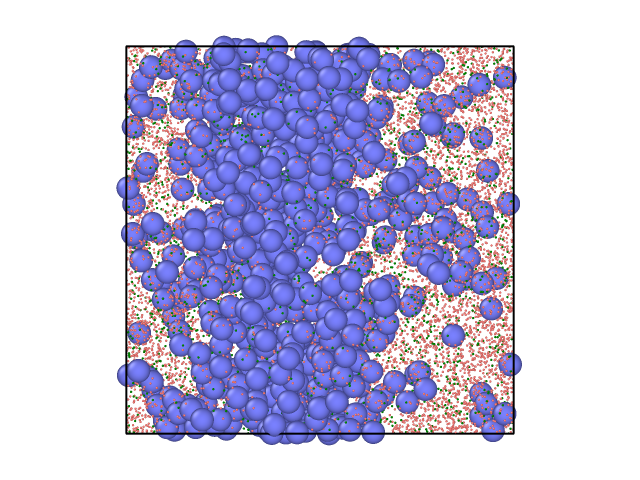}
    \caption{Typical snapshots of one system obtained at stationary state from BD simulations, where enzyme particles are in blue, and $S$ (resp. $P$) particles represented as red (resp. green) dots.  On the left, the density of $S/P$ particles is low, $X=v_\te\rho_{\ts\tp}^0=0.357$, and the system is homogeneous. On the right, the density of $S/P$ particles is high, $X=v_\te\rho_{\ts\tp}^0=1.191$ and the enzyme density is non homogenous. Parameters of the corresponding simulations are:   $\sigma_{\te\ts}=\sigma_{\te\tp}=1.5$, $\rho_\te^0=0.02$, $\varepsilon=-8$, $\mu=16$, $x=k_\text{cat}\rho_\te^0/k_\text{spo}=0.267$, $\alpha=7$.
}
 \label{fig:snapshots}
\end{figure}

\subsection{Analyzing the spatial density of enzymes}

A visual inspection of typical simulation snapshots first reveals whether the enzyme density is homogeneous or not. For instance, Fig. \ref{fig:snapshots} shows two snapshots obtained in a homogeneous case and in an unstable case. To quantitatively characterize the  spatial density of enzymes, we divide the simulation box into cubic sub-boxes of linear size $l$ and compute the local density $\rho_\te(l)$ in each sub-box. The probability distribution $P(\rho_\te)$ of these local densities is then obtained by repeating this process over time, averaging over all sub-boxes, and using a shifting technique by displacing the sub-box grid multiple times, following the method described in Ref.~\cite{Rovere1990}. 

\begin{figure}
    \centering
    \includegraphics[width=1\linewidth]{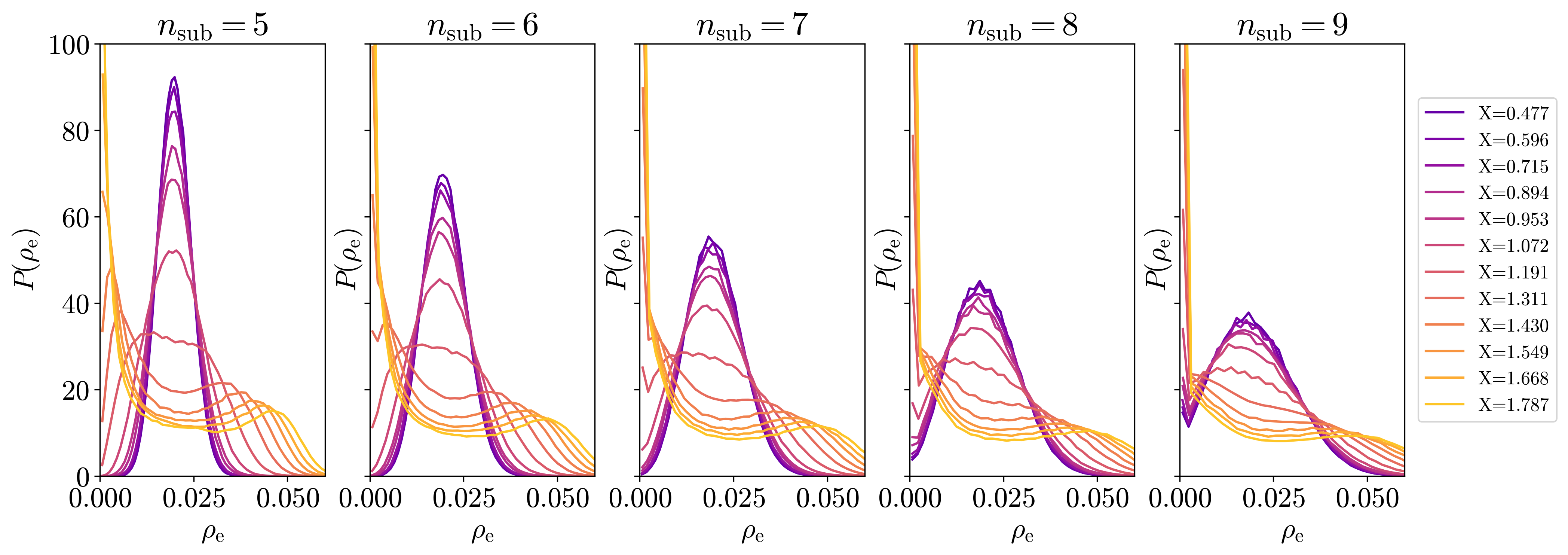}
    \caption{Probability distribution $P(\rho_\te)$ of the local density of enzymes obtained by BD simulations, for $\sigma_{\te\ts}=\sigma_{\te\tp}=1.5$, $\alpha=5$, and $\rho_\te^0=0.02$, for several sizes of sub-boxes, as a function of the density of  substrate/products $X=v_\te\rho_{\ts\tp}^0$. The parameter $n_{\rm sub}$ indicates the number of sub-boxes in each direction, the size of a sub-box being equal to $l=L_{\rm box}/n_{\rm sub}$.
    Other parameters: $\varepsilon=-8$, $\mu=16$, $D_\ts=D_\te$, $x=k_\text{cat}\rho_\te^0/k_\text{spo}=0.267$.}
    \label{fig:supp_p_of_rho}
\end{figure}

\begin{figure}
    \centering
    \includegraphics[width=1\linewidth]{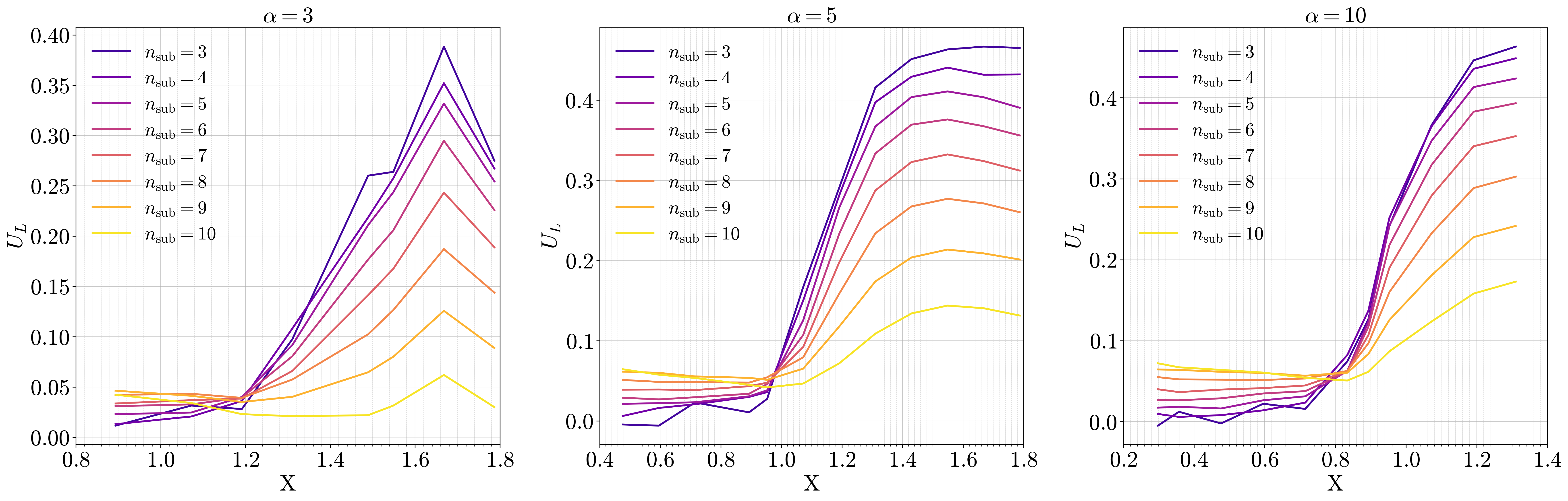}
    \caption{Binder cumulant $U_L$ calculated from the histograms $P(\rho_\te)$ as a function of $X$, for several values of the number of sub-boxes $n_{\rm sub}$, for several values of $\alpha$. The critical value $X_{\text{crit}}$ is identified as the point where the curves for different $n_{\rm sub}$ intersect. Here we have: ($\alpha=3$, $X_{\rm crit}=1.19$),   ($\alpha=5$, $X_{\rm crit}=0.97$),    ($\alpha_p=10$, $X_{\rm crit}=0.79$). Other parameters of BD simulations are  $\sigma_{\te\ts}=\sigma_{\te\tp}=1.5$, , $\rho_\te^0=0.02$, $\varepsilon=-8$, $\mu=16$, $D_\ts=D_\te$, $x=k_\text{cat}\rho_\te^0/k_\text{spo}=0.267$.}
    \label{fig:supp_binder}
\end{figure}

As an example, Fig. \ref{fig:supp_p_of_rho} provides the probability distributions $P(\rho_\te)$ obtained for one system, for several sizes of sub-boxes. As it appears on these plots, it is easy to discriminate between cases where enzymes are homogeneously distributed in the simulation box (typically for $X < 0.715$ in the example of Fig. \ref{fig:supp_p_of_rho}) and cases where the mixture is unstable (typically for $X>1.311$ in the example of Fig. \ref{fig:supp_p_of_rho}). To unambiguously
determine the critical value of the density of  substrate/products $X_{\text{crit}}$ for which phase-separation occurs, we plot the Binder cumulant $U_L$ calculated from the histograms $P(\rho_\te)$ as a function of $X$, for several values of the number of sub-boxes $n_{\rm sub}$. The critical value $X_{\text{crit}}$ is identified as the point where the curves for different $n_{\rm sub}$ intersect, analogous to the approach illustrated in Fig.~15 of Ref.~\cite{Rovere1990}. As an example, Fig. \ref{fig:supp_binder} provides for several systems the Binder cumulant as a function of $X$, for several values of $n_{\rm sub}$. 
To compute $P(\rho_\te)$, we used configurations stored every $4\,\tau_0$,
with $200$ shifts of the sub-box center. Analyzing the time evolution of
these probability distributions, we observed that a stationary state is
reached after a time equal to $8\times 10^3$ in all cases. The Binder cumulant is therefore computed from the probability
distributions averaged over the stationary regime, {\em i.e.} from the
moment the stationary state is reached until the end of the trajectory.

\subsection{Supplemental Movie}

Brownian dynamics of two systems differing only in the substrate/product density: $X = v_{\te} \rho^0_{\ts\tp} = 0.357$ (left, homogeneous) and $X = 1.191$ (right, phase separation). Positions are projected along a given direction; enzymes are shown as semi-transparent blue disks, substrate and product as red and green dots, respectively. Bottom: distribution $P(\rho_{\te})$ of the local enzyme density ($n_\text{sub} = 5$), with the initial distribution as a dashed line. Parameters as in Fig. \ref{fig:snapshots}.

\subsection{Phase diagram at equilibrium}
\begin{figure}
    \centering
    \includegraphics[width=0.8\linewidth]{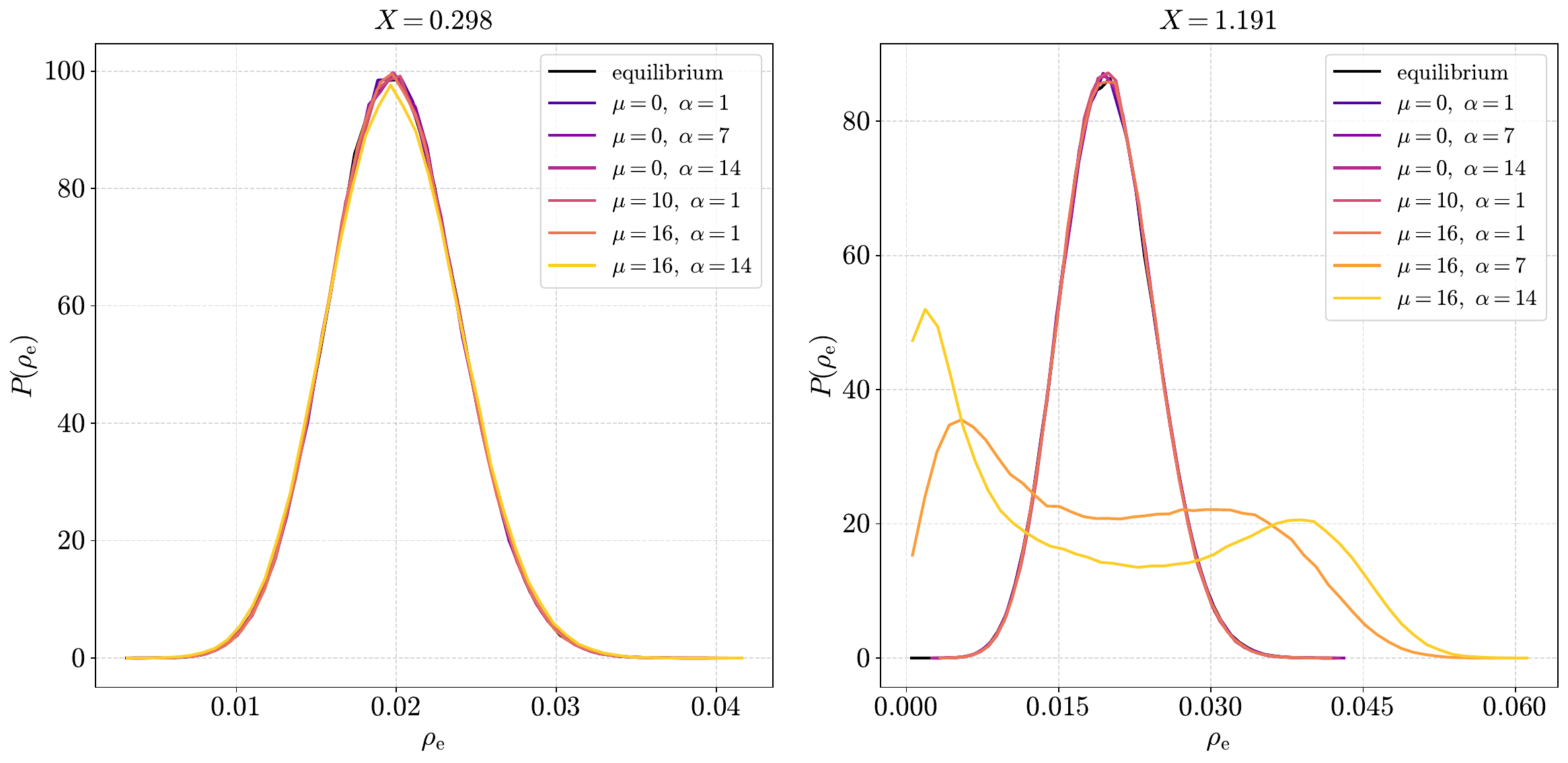}
    \caption{Probability distribution $P(\rho_\te)$ of the local density of enzymes obtained for   $\sigma_{\te\ts}=\sigma_{\te\tp}=1.5$, $\rho_\te^0=0.02$, $\varepsilon=-8$, $x=k_\text{cat}\rho_\te^0/k_\text{spo}=0.267$, at equilibrium and for different values of $\mu$ and $\alpha$. 
}
    \label{fig:supp_equilibrium}
\end{figure}

To obtain the phase diagram of the system at equilibrium, we performed simple Brownian dynamics simulations without any conversions between $S$ and $P$ particles. In the main text, we stated that the equilibrium phase diagram is recovered in the presence of reactions when either $\mu=0$ (regardless of the choice of $D_\ts$ and $D_\tp$), or when $D_\ts=D_\tp$ (regardless of the choice of $\mu$).
To verify this statement, we performed simulations with $\sigma_{\rm es}=\sigma_{\rm ep}=1.5$, $\varepsilon=-8$, and $x=k_{\text{cat}}\rho_{\rm e}^0/k_{\text{spo}}=0.267$, for $5$ different values of $X$.  The probability distribution $P(\rho_\te)$ of the local density of enzymes is indeed the same 
whether equilibrium is achieved: (i) without any reactions, (ii) with reactions and $\mu=0$ for any $\alpha$, or (iii) with reactions and $\alpha=1$ for any $\mu$. As an example we provide the results obtained for $X=0.298$ and $X=1.191$ in Fig. \ref{fig:supp_equilibrium}, together with the results obtained in the presence of reactions, where phase separation occurs. 

\bibliography{refs.bib}